%Plain TeX

\magnification=\magstep1
\font\summ=cmr9
\font\gr=cmbx10 scaled \magstep1
\font\nom=cmcsc10
\font\gt=cmbx12 scaled \magstep2

\centerline {\gt On the Asserted Clash}\vbox to 0.4 cm {}
\centerline {\gt between the Freud and
the Bianchi Identities}\vbox to 1.0 cm {}
\centerline {\nom S. Antoci}\par\vbox to 0.2 cm {}
\centerline {\sl Dipartimento di Fisica ``A.Volta'', Pavia,
Italy}\vbox to 0.5 cm {}
\centerline {\summ (ricevuto il 20 ottobre 1994; approvato il 13
dicembre 1994.)}\vbox to 1.0cm {}

\noindent {\bf Summary.}{\summ - Through a constructive method it
is shown that the claim advanced in recent times about a clash that
should occur between the Freud and the Bianchi identities in
Einstein's general theory of relativity is based on a faulty
argument.\par
}
\par\vbox to 1.0 cm {}

In an article published in this Journal [1] it is asserted that in
Einstein's general theory of relativity the contracted Bianchi
identity and the Freud identity [2] ``clash and lead to a
mathematical overdetermination which creates insurmountable
internal difficulties for the curved-space-time theory of
gravitation as a whole''. The present paper retries the steps of
the argument that led to the quoted assertion, and shows that some
of them are either wrong or not proven.\par We shall deal with
tensorial entities as well as with pseudo-tensorial ones, {\sl
i.e.} with geometric objects that transform tensorially only under
the group of the affine transformations, and eventually with
pseudo-tensorial entities that vanish everywhere in some coordinate
system. The latter objects are called non-tensorial in ref. [1].
The same notation will be used both for the tensorial and for the
pseudo-tensorial entities, in particular boldface letters will be
used to denote both tensor and pseudo-tensor densities.\par Let
$g_{ik}$ be a pseudo-Riemannian metric tensor; ${\bf
g}^{ik}=(-g)^{1/2}g^{ik}$ is the associated contravariant tensor
density, and $\Gamma^i_{km}$ represents the Christoffel symbol
built with $g_{ik}$. The Ricci tensor $R_{ik}(\Gamma)$, the scalar
density ${\bf R}={\bf g}^{ik}R_{ik}$ and the stress-energy-momentum
tensor $T_{ik}$ are then defined, in keeping with the conventions
chosen by Schr\"odinger [3]. We contemplate the pseudo-scalar
density

$${\bf L}={\bf g}^{ik}
(\Gamma^b_{ak}\Gamma^a_{ib}
-\Gamma^b_{ab}\Gamma^a_{ik});\eqno(1)$$
one can write [3]

$${\bf T}^{km}={\delta{\bf R}\over\delta g_{km}}
=-{\delta{\bf L}\over{\delta g_{km}}}
={{\partial\over{\partial x_a}}
\left (\partial{\bf L}\over{\partial g_{km,a}}\right )}
-{\partial{\bf L}\over\partial g_{km}},\eqno(2)$$
where the comma means the ordinary derivative. When ${\bf T}^{km}$
is substituted in it, the contracted Bianchi identity comes to
read:

$${\bf T}^i_{~k;i}={\bf T}^i_{~k,i}
-{1\over 2}{\bf T}^{pq}g_{pq,k}=0,\eqno(3)$$
where the semicolon indicates the covariant derivative done with
respect to $\Gamma^i_{km}$. But then

$${\bf T}^{pq}g_{pq,k}
={{\partial\over{\partial x_a}}
\left ({\partial{\bf L}\over{\partial g_{pq,a}}}g_{pq,k}\right )}
-{\partial{\bf L}\over{\partial x_k}}\eqno(4)$$
and, if we define the pseudo-tensor density of Einstein as

$${\bf u}^i_{~k}
={1\over 2}\left (\delta^i_k{\bf L}
-{\partial{\bf L}\over{\partial g_{pq,i}}}g_{pq,k}\right )
={1\over 2}\left (\delta^i_k{\bf L}
-{\partial{\bf L}\over{\partial {\bf g}^{pq}_{~~,i}}}
{\bf g}^{pq}_{~~,k}\right ),\eqno(5)$$ the contracted Bianchi
identity (3) takes the form

$$\left ({\bf T}^i_{~k}+{\bf u}^i_{~k}\right )_{,i}=0.\eqno(6)$$

These well known developments are recalled here for clearness; we
note in passing that the additional claim of ref. [1], that in
general relativity ${\bf u}^i_{~k,i}=0$, already entails that ${\bf
T}^i_{~k,i}=0$, {\sl i.e.} the very conclusion that is later
reached in that paper through the argument of the clash between the
Bianchi and the Freud identities. This conclusion is contradicted
by a simple example. In fact, if $T_{ik}=\lambda g_{ik}$, where
$\lambda$ is a constant, then

$${\bf T}^i_{~k,i}=\lambda(-g)^{1/2}_{,k},\eqno(7)$$
and the vanishing of the right-hand side of (7) can only be ensured
in particular systems of coordinates.\par

In ref. [1] ${\bf u}^i_{~k}$ is assumed to split into a tensor
density and some entity that can be made to vanish everywhere in
some system of coordinates. Let us provide a construction that
actually performs such a splitting. We first recall [4] that

$${\partial{\bf L}\over{\partial {\bf g}^{pq}_{~~,i}}}
=-\Gamma^i_{pq}+{1\over 2}\delta^i_q\Gamma^a_{pa}
+{1\over 2}\delta^i_p\Gamma^a_{qa}\eqno(8)$$
and since ${\bf
g}^{pq}_{~~;k}=0$, one readily gets

$$\eqalign{{\bf u}^i_{~k}
&={1\over 2}\Big[\delta^i_k{\bf g}^{pq}
(\Gamma^b_{aq}\Gamma^a_{pb}-\Gamma^b_{ab}\Gamma^a_{pq})\cr
&-(\Gamma^i_{pq}-{1\over 2}\delta^i_q\Gamma^a_{pa}
-{1\over 2}\delta^i_p\Gamma^a_{qa})
\cdot({\bf g}^{bq}\Gamma^p_{bk}+{\bf g}^{bp}\Gamma^q_{bk}
-{\bf g}^{pq}\Gamma^b_{kb})\Big].\cr }
\eqno \hbox{$(9)$} $$
 We register that

$${\bf u}^i_{~i}={\bf g}^{ik}(\Gamma^b_{ak}\Gamma^a_{ib}
-\Gamma^b_{ab}\Gamma^a_{ik})={\bf L}.\eqno(10)$$
 Through eq. (9) a constructive approach to
the splitting of ${\bf u}^i_{~k}$ in the above-mentioned way
becomes apparent. Besides the pseudo-Riemannian metric $g_{ik}$,
let us consider a Minkowskian metric $s_{ik}$, whose Christoffel
symbols $\Sigma^i_{km}$ can be transformed into zero everywhere in
some coordinate system. One writes [5]

$$\Gamma^i_{km}=\Sigma^i_{km}+\Delta^i_{km},~~
\Delta^i_{km}={1\over 2}g^{ip}
(g_{pk|m}+g_{pm|k}-g_{km|p}),\eqno(11)$$
 where ``$|$'' indicates the covariant derivative done with respect to
$\Sigma^i_{km}$. Therefore, $\Delta^i_{km}$ is a tensor, but we
shall not leave unnoticed that its definition in by no means unique
since, for a given $g_{ik}$, the Minkowskian metric $s_{ik}$ can be
so chosen as to ensure the overall vanishing of the connection
$\Sigma^i_{km}$ in whatever coordinate system one likes, and the
tensor $\Delta^i_{km}$ extracted from $\Gamma^i_{km}$ will depend
on that choice. By substituting the $\Gamma^i_{km}$ given by eq.
(11) in eq. (9) the latter can be rewritten as

$${\bf u}^i_{~k}={\bf t}^i_{~k}+{\bf w}^i_{~k},\eqno(12)$$
 {\sl i.e.} as the sum of a non-uniquely defined tensor density
${\bf t}^i_{~k}$ and of ${\bf w}^i_{~k}$, which by construction
vanishes in the coordinate system in which $\Sigma^i_{km}$ is
everywhere zero. Hence in that system ${\bf t}^i_{~i}={\bf
u}^i_{~i}={\bf L}$. But, since in that coordinate system the metric
tensor $g_{ik}$ can be so chosen as to ensure that ${\bf L}$ does
not vanish, it turns out that the scalar density ${\bf t}^i_{~i}$
does not necessarily vanish, and the same occurrence happens to
${\bf t}^i_{~k}$. The assertion to the contrary contained in ref.
[1] is thereby disproved.\par

Let us define now the pseudo-tensor density

$${\bf U}^i_{~k}={\bf T}^i_{~k}+{\bf u}^i_{~k}.\eqno(13)$$
 Freud has shown [2] that one can write

$${\bf U}^i_{~k}={\partial{\bf V}^{ia}_k
\over{\partial x_a}},\eqno(14)$$
 where the pseudo-tensor density ${\bf V}^{ia}_k$ is skew in the
 upper indices and can be defined by the expansion of a determinant:

$${\bf V}^{ia}_k={1\over 2}\left\vert
\matrix{\delta^i_k&\delta^a_k&\delta^b_k \cr
{\bf g}^{is}&{\bf g}^{as}&{\bf g}^{bs} \cr
\Gamma^i_{bs}&\Gamma^a_{bs}&\Gamma^b_{bs} \cr}
\right\vert .\eqno(15)$$
 We introduce now another Minkowskian metric tensor, $l_{ik}$, whose
Christoffel symbols $\Lambda^i_{km}$ will vanish in a coordinate
system that in general differs from the one in which the
$\Sigma^i_{km}$ vanish. We can write

$$\Gamma^i_{km}=\Lambda^i_{km}+\Theta^i_{km},~~
\Theta^i_{km}={1\over 2}g^{ip}
(g_{pk:m}+g_{pm:k}-g_{km:p}),\eqno(16)$$
 where $\Theta^i_{km}$ is a tensor built with the covariant
derivative ``$:$'' done with respect to $\Lambda^i_{km}$, and is
not uniquely defined, due to the freedom of choice for $l_{ik}$. An
elementary property of determinants ensures that, by substituting
(16) in (15), one can write

$${\bf V}^{ia}_k={\bf X}^{ia}_k+{\bf Z}^{ia}_k,\eqno(17)$$
{\sl i.e.} one can split ${\bf V}^{ia}_k$ into a skew tensor
density ${\bf X}^{ia}_k$ and a skew object ${\bf Z}^{ia}_k$ that
vanishes when the $\Lambda^i_{km}$ are zero. The splitting is once
more not unique: it depends on the choice of $l_{ik}$. We aim at
achieving the same sort of splitting for ${\bf U}^i_{~k}$, {\sl
i.e.} at writing:

$${\bf U}^i_{~k}={\bf x}^i_{~k}+{\bf z}^i_{~k},\eqno(18)$$
where ${\bf x}^i_{~k}$ is a tensor density,
while ${\bf z}^i_{~k}$ is zero in the coordinate system in which
$\Lambda^i_{km}$ vanishes. Since

$${\bf X}^{ia}_{k,a}={\bf X}^{ia}_{k:a}
+{\bf X}^{ia}_n\Lambda^n_{ka}\eqno(19)$$
 is not a tensor density, we cannot pose ${\bf x}^i_{~k}={\bf
X}^{ia}_{k~,a}$. A position that fulfils our requirements is:

$${\bf x}^i_k={\bf X}^{ia}_{k:a},~~{\bf z}^i_k
={\bf X}^{ia}_n\Lambda^n_{ka}+{\bf Z}^{ia}_{k,a},\eqno(20)$$
 but, of course, it does not allow to conclude that ${\bf
x}^i_{~k,i}$ and ${\bf z}^i_{~k,i}$ must vanish separately due to
the play of the indices.\par The constructive method leads
eventually to rewrite eq. (13) as follows:

$${\bf T}^i_{~k}={\bf U}^i_{~k}-{\bf u}^i_{~k}
={\bf x}^i_{~k}-{\bf t}^i_{~k}
+{\bf z}^i_{~k}-{\bf w}^i_{~k},\eqno(21)$$
 where ${\bf z}^i_{~k}-{\bf w}^i_{~k}$ is a non-vanishing tensor
density, that can be annihilated through the particular choice
$s_{ik}=l_{ik}$ for the two Minkowskian metrics. But this choice is
by no means mandatory; therefore ${\bf U}^i_{~k}-{\bf u}^i_{~k}$
shall not be generally split into two tensorial parts only, as
asserted in ref. [1].\par
     In conclusion: a constructive method allows to appreciate
that the splitting of ${\bf U}^i_{~k}-{\bf u}^i_{~k}$ into
tensorial parts is arbitrary and generally threefold, that ${\bf
t}^i_{~k}$ is not generally vanishing and that ${\bf x}^i_{~k,i}$
is not necessarily zero due to symmetry reasons. The argument
offered in ref. [1] for proving that in general relativity a clash
between the Freud and the Bianchi identities leads to the vanishing
of ${\bf T}^i_{~k,i}$ requires instead a twofold splitting of ${\bf
U}^i_{~k}-{\bf u}^i_{~k}$, a vanishing ${\bf t}^i_{~k}$ and a zero
${\bf x}^i_{~k,i}$. That argument therefore avails of either wrong
or unproven premises.\par
\centerline {* * *}\par\noindent
I express my gratitude to E. Giannetto for addressing my attention
to this question.\par\vbox to 1.2cm {}

\noindent{\gr References}\par\vbox to 0.8cm {}
\noindent [1]~~Yilmaz H., {\sl Nuovo Cimento B}, {\bf 107}, (1992)
941.\par
\noindent [2]~~Freud Ph., {\sl Ann. Math.}, {\bf 40}, (1939) 417.\par
\noindent [3]~~Schr\"odinger E., {\sl Space-Time Structure} (Cambridge
University  Press, Cambridge 1954).\par
\noindent [4]~~Schr\"odinger E., {\sl Space-Time Structure} (Cambridge
University Press, Cambridge 1954), p. 103.\par
\noindent [5]~~Rosen, N., {\sl Phys. Rev.}, {\bf 57}, (1940) 147.
\end